*Type of article*

# Non-Ideal Two-Level Battery Charger—Modeling and Simulation

**José M. Campos-Salazar[1,*], Juan L. Aguayo-Lazcano [2], and Roya Rafiezadeh [3]**

[1] Electronic Engineering Department, Electronic Engineering Department, Barcelona, Spain
[2] Institute of Physical and Mathematical Sciences, Universidad Austral de Chile, Valdivia, Chile
[3] PEMC group, University of Nottingham, Nottingham, United Kingdom

* **Correspondence:** jose.manuel.campos@upc.edu.

**Abstract:** This study presents a comprehensive analysis of a two-level battery charger for electric vehicles, focusing on modeling, simulation, and performance evaluation. The proposed charger topology employs two switches operating complementarily, along with essential components such as inductors, capacitors, and batteries. Detailed modeling in steady-state and dynamic regimes reveals the influence of nonlinearities, particularly switch and energy storage element characteristics, on charger efficiency and performance. Efficiency calculations highlight the significance of precise modeling and control strategies. Controller synthesis involves designing a robust proportional-integral compensator for effective battery charging current regulation. Analysis of non-idealities underscores the need for accurate component sizing and control strategies. The findings provide valuable insights for optimizing charger design and control, with implications for enhancing electric vehicle performance and reliability.

## 1. Introduction

The surge in electric vehicle (EV) adoption is largely due to growing concerns about climate change, driven by unsustainable greenhouse gas emissions from fossil fuel-powered transportation [1,2]. Collaborative initiatives between government agencies and the private sector aim to move away from non-renewable energy sources, with significant investments being made to make EVs more competitive with conventional vehicles.

At the heart of EVs is the battery, a critical component that determines key metrics such as range, performance and reliability. Consisting of battery cells organized into modules that are further interconnected to form battery packs, its effectiveness is monitored by the battery management system (BMS) [3]. The BMS monitors electrical and thermal parameters, cell charge balance, and auxiliary functions to ensure safety and efficiency.

In addition to batteries, EV efficiency depends heavily on power converters, particularly the traction inverter and battery charger (BC). While the former manages the exchange of power between the battery and the engine, the latter facilitates the transfer of power between the battery and the grid or alternative sources. This study focuses on battery charging systems, recognizing their critical role in optimizing key vehicle attributes such as charging speed, range, battery life, and potential grid interaction [4].

Battery chargers are categorized based on several factors: conductive vs. inductive charging, on-board (OBC) vs. off-board (FBC) placement, charging speed, and unidirectional vs. bidirectional power flow. Standards such as IEC 61851-1 and SAE J1772 further classify chargers based on charging mode, type and level [1].

BCs can be broadly classified according to their topologies in the case of OBCs and FCBs., focusing on power supply, converter technology, and the presence of galvanic isolation. OBCs typically include an EMI filter, an ac-dc converter with PFC, a dc link, and a dc-dc converter, with variants allowing bi-directional power flow [2].

For FBCs, notable topologies include single-phase and three-phase isolated and non-isolated dc-dc converters, each with different efficiencies, power densities, and challenges [2]. Modulation schemes such as PSM and DPWM optimize performance, while designs such as the three-level-boost converter promise low harmonic distortion and bi-directional power flow [5,6].

In-depth analyses of OBC and FBC configurations highlight the benefits of multilevel technology, despite challenges in system modeling and control optimization. Future research must address these gaps to advance the understanding and optimization of fast battery charging technology [7,8].

This study will focus on the analysis of a two-level dc-dc BC (2L-BC). There is a substantial body of literature on 2L-BC, with numerous studies and articles published on the subject. For instance, the work in [9] first introduces the topic of dc-dc converters for mobile battery energy storage system (BESS) charging EVs with dc fast-charger (DCFC) and outlines its objectives and structure. It then delves into the background information necessary to understand the research context, covering relevant standards, safety requirements and technical specifications. A comprehensive review of the existing literature on dc-dc converters and related technologies is provided to establish a foundation for the research. The methodology for selecting and simulating dc-dc converter topologies is detailed, with a focus on buck-boost and dual-active-bridge converters. Simulation results are presented and analyzed, comparing efficiency and performance between different transistor types (IGBTs vs. SiC MOSFETs) and discussing the implications of the results. The thesis concludes with key lessons learned from the research, highlighting the most promising converter topologies and outlining potential areas for future research and development. Overall, it provides a thorough exploration of dc-dc converter design for mobile BESS charging EVs with DCFCs, combining theoretical background, simulation-based analysis, and practical implications for real-world implementation.

A review and survey of future trends of power converters for fast-charging stations of electric vehicles is presented in [10]. The introduction provided an overview of the diverse landscape of EVs, encompassing battery electric vehicles, plug-in hybrid electric vehicles, and hybrid electric vehicles. It also highlighted the environmental benefits of EVs, including reduced greenhouse gas emissions. Challenges associated with EVs, such as limited driving range and the need for an extensive charging infrastructure, were acknowledged. The discussion then turned to the topic of EV charging infrastructure. This included an examination of ac charging and dc fast-charging methods, which were categorized according to their respective power delivery levels. The significance of standards such as SAE J1772 and CCS for ac and dc charging, respectively, was also highlighted. In the course of examining DCFC, the transition from rapid to ultra-fast charging schemes was elucidated, accompanied by a delineation of power levels and connector types, including CHAdeMO, GB/T, and Type 1/Type 2 CCS. The classification of DCFC conversion stages introduced various rectifier topologies, including the diode bridge, Vienna rectifier, PWM rectifier, and NPC rectifier, as well as isolated and non-isolated dc-dc converter topologies, such as LLC, DAB, and phase-shift full-bridge converters. In examining future research trends, the focus was on modular multilevel converter (MMC) topologies for high-power applications such as DCFC. This approach emphasized the advantages of these topologies in terms of fault tolerance, scalability, and efficiency. Additionally, the modular push-pull converter (MPC) was introduced as a promising alternative topology. In conclusion, the paper summarized its findings, encapsulating an overview of EV charging infrastructure, a classification of charger conversion stages, and future research directions. It also underscored the pivotal role of MMCs and MPCs in advancing DCFC technology.

The article [11], presents an in-depth exploration of an offline battery charger tailored for plug-in EVs, leveraging a buck-boost power factor correction (PFC) converter. The pivotal role of batteries within the context of EVs is highlighted, with particular emphasis on factors such as cost, energy density, and charging time. The distinction between on-board and off-board charging systems is delineated, with particular emphasis on the latter's potential for high charging power levels and reduced size constraints. Previous works integrating battery chargers into electrical drive systems are referenced, thus setting the stage for the proposed off-line battery charger's introduction. The novel converter modifies the traditional three-phase voltage source inverter (VSI) and motor drive into a battery charger system, employing a buck-boost topology with PFC control. In operation, the converter alternates between buck and boost modes to facilitate efficient charging, with the current being shaped by a PFC controller to match the line frequency. The simulation results validate the proposed charger's performance, confirming its suitability for plug-in EVs through high input power factor and robust charging capabilities. The paper concludes by affirming the versatility and reliability of the proposed system, suggesting its applicability across diverse battery sizes and high-power EV applications.

A survey on non-isolated high-voltage step-up dc–dc topologies based on the boost converter is shown by [12], the literature review offers a comprehensive examination of boost-based dc-dc converter topologies, with a particular focus on their suitability for high-voltage step-up applications. It is evident that conventional boost converters are inherently limited in their ability to effectively handle high-power levels. This is due to their reliance on increasing duty cycles, which ultimately leads to increased losses and reduced efficiency. This reliance not only leads to increased losses and reduced efficiency but also necessitates complex and costly drive circuitry. To address these challenges, alternative topologies have been explored. Three-level converters show promise by doubling the static gain compared to conventional boost converters, resulting in advantages such as reduced size, weight, and volume of filter inductors. However, they may not achieve as wide a conversion ratio as cascaded converters and can experience significant voltage stress across the active switch. Hybrid approaches, such as quadratic three-level boost converters, offer potential solutions but may be limited by the presence of bulky inductors in high-power applications. Coupled inductors provide a simple solution for high-voltage step-up by adjusting the turns ratio, but may suffer from voltage ringing and poor efficiency due to resonance effects. Switched capacitors offer a modular approach to achieving high-output voltages but require careful consideration of trade-offs between component count, efficiency, and static gain. Interleaved converters and topologies based on the third switched-source cell (3SSC) offer solutions for high-power, high-current applications with high-voltage step-up. While interleaved converters require complex control schemes for current sharing, 3SSC-based topologies naturally maintain current sharing due to the existence of an autotransformer with unity turns ratio. Recent research indicates the potential of 3SSC-based topologies for novel high-voltage-step converter designs with high efficiency. This suggests promising avenues for further exploration and development in the field. Overall, the conclusion underscores the importance of exploring alternative topologies beyond conventional boost converters to address the challenges of achieving high-voltage step-up in high-power applications while optimizing efficiency and performance.

From the aforementioned works, it can be observed that none of the converters included in the stages of the BCs have considered the nonlinearities present in the switches and energy storage elements. However, there are some studies in the literature that have examined dc-dc converters and incorporated such nonlinearities, although they are not directly related to BCs. For instance, the article [13] presents a comprehensive examination of the intricacies of modeling non-ideal dc-dc converters, with a particular emphasis on the boost, buck, and buck-boost converter topologies. The article opens by elucidating the challenges posed by switches in power electronics due to their non-linear and time-varying nature. The paper employs an exact state-space modeling approach to derive precise equations for the boost, buck, and buck-boost converters, taking into account the effects of switching on system stability, dc gain, and efficiency. Notably, the study finds that the properties of the buck and boost converters are special cases of the more versatile buck-boost converter. The impact of switching effects, including time modulation and duty cycle variation, on stability and performance metrics is comprehensively explored. The analysis reveals that while the buck converter is relatively robust against switching effects, the boost and buck-boost converters exhibit weaknesses, particularly as the duty cycle approaches unity. Moreover, it is demonstrated that efficiency is optimized when dc gain is close to unity, which leads to recommendations regarding optimal duty factors for each converter type. In conclusion, the study offers valuable insights into the intricacies of non-ideal dc-dc converter modeling, providing essential guidance for high-frequency and high-power applications.

An study regarding of a dc-dc buck-boost converter with non-linear power inductor operating in saturation region considering electrical losses is shown in [14]. The study begins by emphasizing the significance of dc-dc buck-boost power converters in a variety of applications and introduces the goal of developing a control strategy for regulating these converters under non-ideal conditions. A mathematical model of a non-ideal dc-dc buck-boost power converter is developed using the Euler–Lagrange formalism, with a particular focus on electrical losses and inductor saturation. Subsequently, an analysis of the dynamic and stationary behaviors of the power converter is conducted, which provides insights into operational restrictions and controllability, which are crucial for the effective design of an effective control strategy. A linear controller is then designed based on a tuning methodology that leverages a linear approximation of the converter's behavior and frequency domain analysis. The controller design incorporates integral action, lead compensation, and an anti-windup scheme, thereby ensuring robust performance. Further details on the tuning methodology and controller parameters selection are provided, emphasizing robustness, appropriate bandwidth, and improved response time. A comparison is conducted with alternative control strategies, including non-linear PI and adaptive passivity control combined with a linear PI controller, in order to evaluate performance under various operational conditions. The study concludes by summarizing its key findings and highlighting the effectiveness of the proposed linear controller in regulating the non-ideal power converter and its significance in advancing control strategies for dc-dc buck-boost power converters. In conclusion, the study offers a comprehensive exploration of modeling, analysis, and control design for non-ideal dc-dc buck-boost power converters, providing valuable insights for researchers and engineers in the field.

Finally, the work [15] provides a comprehensive examination of the intricacies of non-ideal dc-dc PWM boost converters, encompassing foundational concepts and advanced analysis, as well as experimental validation. Initially, the introduction sets the stage by elucidating the importance of addressing non-idealities in converter design for accurate performance characterization. It proceeds to elucidate the operational principles of boost converters, elucidating the functions of key components such as switches, inductors, capacitors, and diodes. Subsequently, a detailed analysis of non-idealities ensues, meticulously examining their impact on inductor ripple current and capacitor voltage ripple. This analysis lays the groundwork for deriving precise equations for inductor and capacitor design, factoring in non-ideal elements to ensure optimal performance. Furthermore, the article scrutinizes the implications of equivalent series resistance (ESR), elucidating its critical role in maintaining output voltage stability within predefined thresholds. Subsequently, small signal analysis is conducted to derive transfer function models that facilitate dynamic performance evaluation, which is crucial for controller design.

Simulation and experimental results are presented to corroborate theoretical analyses, demonstrating the tangible effects of non-idealities on converter behavior and the efficacy of proposed design methodologies. The conclusion underscores the significance of meticulous modeling and design techniques for non-ideal boost converters, emphasizing their pivotal role in controller design and real-world applications. This study advocates for the adoption of the presented non-ideal transfer function model to inform robust controller design and suggests avenues for future research. In essence, the paper provides a comprehensive understanding of non-idealities in boost converters, offering valuable insights for engineers and researchers in the field of power electronics.

The objective of this proposed paper is to present a battery charger based on a two-level dc-dc power converter. The focus will be on the non-linearity related to the switches and energy storage elements. The modeling stage is systematically developed, resulting in the switching model, which is then averaged and analyzed in small signal. Subsequently, a feedback output compensator is derived with the purpose of regulating the battery charging current. The effects of the non-linarites on the steady state of the battery charger and the behavior of the energy storage elements are then analyzed.

The following section outlines the structure of the paper. Sections 2 and 3 present the proposed topology and develop the modeling of the battery charger. Subsequently, Sections 4 and 5 derive the nonlinear expression of the charger performance-dependent gain and the equations related to the sizing of the charger inductor and capacitor, based on the inductor current and capacitor voltage ripple, respectively. In Sections 6 and 7, a feedback output compensator that controls the charger behavior is synthesized, and the simulation results of the system are presented. Finally, the conclusions drawn from the work are presented in Section 8.

## 2. Battery charger topology

In **Figure 1**, the proposed topology of the battery charger (2L-BC) is shown. Basically, it is based on a two-level dc-dc converter, and its load is modeled as a standard battery [16]. The 2L-BC is configured by two switches named $Q_1$ and $Q_2$ that operate complementarily [17]. Each switch is implemented using a MOSFET device including its $R_{DS(on)}$ driving resistors. In addition, the 2L-BC incorporates the series resistors of $L$ and $C$, i.e., $r_L$ and $r_C$ respectively. Finally, the battery is modeled using an internal voltage $v_{OB}$ in series with its internal resistor $r_B$. Regarding the system variables, $v_d(t)$, $v_L(t)$ and $v_C(t)$ are the input, $L$ and $C$ voltages, respectively. On the other hand, $i_L(t)$, $i_C(t)$ and $i_B(t)$ are the currents of $L$, $C$ and the battery, respectively. Finally, it is assumed that the 2L-BC operates in continuous conduction mode.

## 3. Battery charger modeling

This section presents the modeling of the charger in both steady and dynamic regimes.

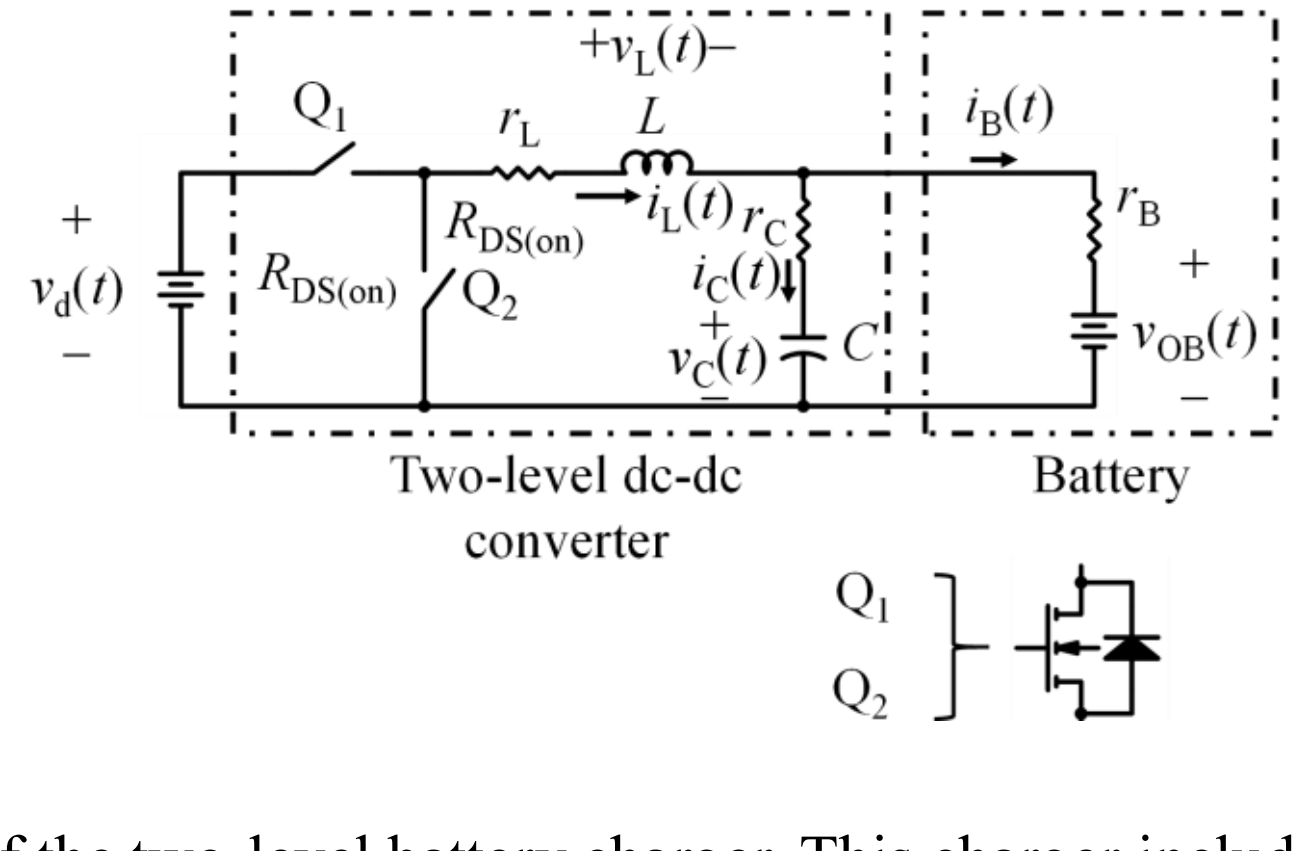

**Figure 1.** Topology of the two-level battery charger. This charger includes the internal resistors of the MOSFET devices ($R_{DS(on)}$), and the series resistors of $L$ ($r_L$) and $C$ ($r_C$).

### *3.1. Modeling 2L-BC in steady state*

As a preliminary step in the modeling process, it is essential to identify the switching strategy employed by the converter switches (Q1 and Q2). This enables the determination of the switching function of the system in a steady state, which is represented by the variable $s_f(t)$. The switching frequency of the 2L-BC, is labeled as $f_s$, and the switching period is defined as $T_s = 1/f_s$. It is known that both switches operate in a complementary manner, and thus, the following strategy is derived and defined as follows: in the first half cycle of the switching period, i.e., when $0 \leqslant t < D \cdot T_s$, switches $Q_1$ and $Q_2$ operate as closed and open, respectively. Subsequently, for the second half cycle, that is, when $D \cdot T_s \leqslant t < T_s$, $Q_1$ and $Q_2$ operate as open and closed, respectively. With this information it is possible to define $s_f(t)$ as follows [17]:

$$s_f(t) = \begin{cases} 1, & 0 \leq t < D \cdot T_s \\ 0, & D \cdot T_s \leq t < T_s \end{cases} \tag{1}$$

It should be noted that D is the duty cycle under stationary conditions.

The governing equations for the voltage of $L$ ($v_L(t)$) and the current of $C$ ($i_C(t)$) are derived by applying Kirchhoff's laws of voltage and current. These equations are functions of the variation of the topology, as the switching half-cycles occur. The topology in the first switching half-cycle is shown in **Figure 2(a)**, and the expressions for $v_L(t)$ and $i_C(t)$ are given by:

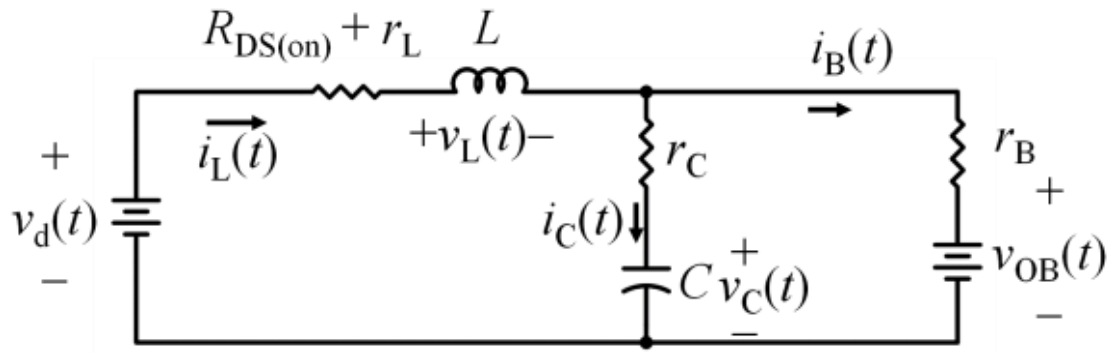

(a)

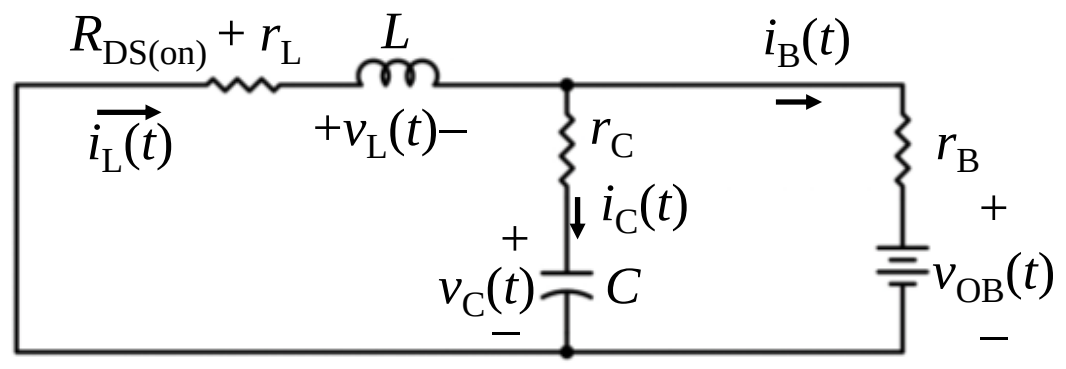

(b)

**Figure 2.** Topologies of 2L-BC as a function of the switching half-cycles. (a) Topology of 2L-BC when $0 \leq t < D \cdot T_s$ ($Q_1$ → closed and $Q_2$ → open). (b) Topology of 2L-BC when $D \cdot T_s \leq t < T_s$ ($Q_1$ → open and $Q_2$ → closed).

$$v_L(t) = v_d(t) - \left(\frac{r_B \,||\, r_C}{r_C}\right) \cdot v_C(t) - \left(\frac{r_B \,||\, r_C}{r_B}\right) \cdot v_{OB}(t) - (R_{in} + r_B \,||\, r_C) \cdot i_L(t) \tag{2}$$

$$i_C(t) = \left(\frac{r_B \,||\, r_C}{r_C}\right) \cdot i_L(t) - \left(\frac{r_B \,||\, r_C}{r_B \cdot r_C}\right) \cdot v_C(t) + \left(\frac{r_B \,||\, r_C}{r_B \cdot r_C}\right) \cdot v_{OB}(t) \tag{3}$$

where $R_{in} = R_{DS(on)} + r_L$. In contrast, the small ripple approximation (SRA) principle is applied to (2) and (3), assuming that the ripple of the voltages $v_d(t)$, $v_C(t)$ and $v_{OB}(t)$ and of the current $i_L(t)$ is sufficiently small to be neglected, a common assumption in practice [17,18]. The revised versions of (2) and (3) following the application of the SRA are presented as follows:

$$v_L(t) \approx V_d - \left(\frac{r_B \,||\, r_C}{r_C}\right) \cdot V_C - \left(\frac{r_B \,||\, r_C}{r_B}\right) \cdot V_{OB} - (R_{in} + r_B \,||\, r_C) \cdot I_L \tag{4}$$

$$i_C(t) \approx \left(\frac{r_B \,||\, r_C}{r_C}\right) \cdot I_L - \left(\frac{r_B \,||\, r_C}{r_B \cdot r_C}\right) \cdot V_C + \left(\frac{r_B \,||\, r_C}{r_B \cdot r_C}\right) \cdot V_{OB} \tag{5}$$

The topology of the system when operating in the second half-cycle is depicted in **Figure 2(b)**, and the expressions of $v_L(t)$ and $i_C(t)$ are defined as follows

$$v_L(t) = -\left(\frac{r_B \,||\, r_C}{r_C}\right) \cdot v_C(t) - \left(\frac{r_B \,||\, r_C}{r_B}\right) \cdot v_{OB}(t) - (R_{in} + r_B \,||\, r_C) \cdot i_L(t) \tag{6}$$

$$i_C(t) = \left(\frac{r_B \,||\, r_C}{r_C}\right) \cdot i_L(t) - \left(\frac{r_B \,||\, r_C}{r_B \cdot r_C}\right) \cdot v_C(t) + \left(\frac{r_B \,||\, r_C}{r_B \cdot r_C}\right) \cdot v_{OB}(t) \tag{7}$$

Then, applying the SRA to (6) and (7), the new equations related to this switching half-cycle are given by:

$$v_L(t) \approx -\left(\frac{r_B \,||\, r_C}{r_C}\right) \cdot v_C(t) - \left(\frac{r_B \,||\, r_C}{r_B}\right) \cdot v_{OB}(t) - (R_{in} + r_B \,||\, r_C) \cdot i_L(t) \tag{8}$$

$$i_C(t) \approx \left(\frac{r_B \,||\, r_C}{r_C}\right) \cdot i_L(t) - \left(\frac{r_B \,||\, r_C}{r_B \cdot r_C}\right) \cdot v_C(t) + \left(\frac{r_B \,||\, r_C}{r_B \cdot r_C}\right) \cdot v_{OB}(t) \tag{9}$$

In light of the fact that the 2L-BC operates in a state of steady-state, the principles of inductor volt-second balance and capacitor charge balance are applied to the system. With regard to the former, it is demonstrated that the average voltage of $L$ over one switching period ($T_s$), i.e. $\langle v_L(t) \rangle_{T_s}$, must be zero [17]. Consequently, the inductor volt-second balance is calculated from (4) and (8), as stated as follows:

$$D \cdot V_d - \left(\frac{r_B \,||\, r_C}{r_C}\right) \cdot V_C - \left(\frac{r_B \,||\, r_C}{r_B}\right) \cdot V_{OB} - \left(R_{in} + r_B \,||\, r_C\right) \cdot I_L = 0 \tag{10}$$

Conversely, the capacitor charge balance principle is applied according to (5) and (9), resulting in the verification that equation $\langle i_C(t) \rangle_{T_s} = 0$, as given by [17]:

$$r_B \cdot I_L - V_C + V_{OB} = 0 \tag{11}$$

It should be noted that (10) and (11) represent the model of the charger in equilibrium (steady state). For this system, the unknown variables are $V_C$ and $I_L$, while the known variables are $D$ and $V_{OB}$. Solving for $V_C$ and $I_L$ yields the steady state variables given by:

$$\begin{cases} I_L = \dfrac{1}{r_B} \cdot \left(V_C - V_{OB}\right) \\ V_C = r_B \,||\, R_{in} \cdot \left(\dfrac{D}{R_{in}} \cdot V_d + \dfrac{1}{r_B} \cdot V_{OB}\right) \end{cases} \tag{12}$$

From (11), it can be observed that the $V_C$ dynamics under steady-state conditions exhibit a slight dependence on the charger nonlinearities, particularly with regard to $R_{DS(on)}$ and $r_L$. This is due to the fact that, in practice, $r_B$ is significantly larger than $R_{in}$, resulting in $r_B \,||\, R_{in} \approx r_B$. Therefore, the equilibrium state of the system may change if such resistances are not considered. The model of the charger in steady state is illustrated in **Figure 3**.

*3.2. Modeling 2L-BC in steady state*

**Figure 2** and (1)–(3), (8) and (9) provide the 2L-BC switching model defined by (13). Subsequently, the averaging operator $\langle x(t) \rangle_{T_s} = \frac{1}{T_s} \cdot \int_{t-T_s}^{t} x(\tau) \cdot d\tau$ is applied to (13) over a $T_s$, resulting

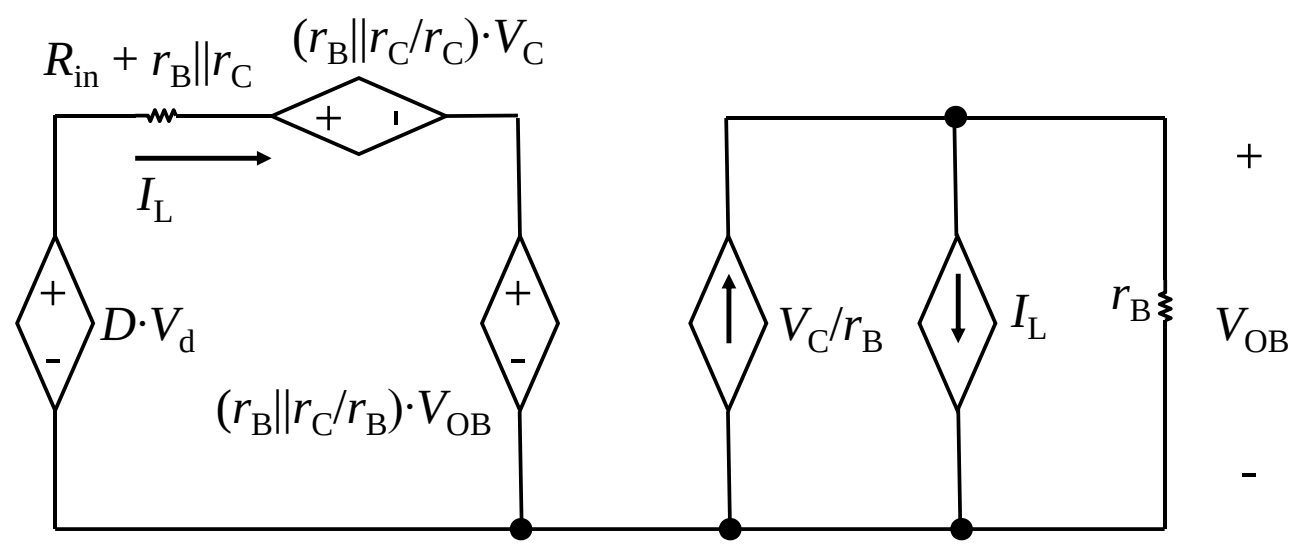

**Figure 3.** Battery-charger steady-state model.

$$\frac{d}{dt}\begin{bmatrix} i_L(t) \\ v_C(t) \end{bmatrix} = \begin{bmatrix} -\frac{R_{in} + r_B || r_C}{L} & -\frac{r_B || r_C}{r_C \cdot L} \\ \frac{r_B || r_C}{r_C \cdot C} & \frac{r_B || r_C}{r_C \cdot r_B \cdot C} \end{bmatrix} \cdot \begin{bmatrix} i_L(t) \\ v_C(t) \end{bmatrix} + $$

$$+\begin{bmatrix} \frac{1}{L}\cdot s_f(t) & -\frac{r_B || r_C}{r_B \cdot L} \\ 0 & \frac{r_B || r_C}{r_C \cdot r_B \cdot C} \end{bmatrix} \cdot \begin{bmatrix} v_d(t) \\ v_{OB}(t) \end{bmatrix} \tag{13}$$

$$i_B(t) = \begin{bmatrix} \left(1 - \frac{r_B || r_C}{r_C \cdot C}\right) & \frac{1}{r_B} \end{bmatrix} \cdot \begin{bmatrix} i_L(t) \\ v_C(t) \end{bmatrix} - \frac{1}{r_B}\cdot v_{OB}(t)$$

in the averaged model of the 2L-BC, defined as following [17,19]:

$$\frac{d}{dt}\begin{bmatrix} \langle i_L(t)\rangle_{T_s} \\ \langle v_C(t)\rangle_{T_s} \end{bmatrix} = \begin{bmatrix} -\frac{R_{in} + r_B || r_C}{L} & -\frac{r_B || r_C}{r_C \cdot L} \\ \frac{r_B || r_C}{r_C \cdot C} & \frac{r_B || r_C}{r_C \cdot r_B \cdot C} \end{bmatrix} \cdot \begin{bmatrix} \langle i_L(t)\rangle_{T_s} \\ \langle v_C(t)\rangle_{T_s} \end{bmatrix} + $$

$$+\begin{bmatrix} \frac{1}{L}\cdot d(t) & -\frac{r_B || r_C}{r_B \cdot L} \\ 0 & \frac{r_B || r_C}{r_C \cdot r_B \cdot C} \end{bmatrix} \cdot \begin{bmatrix} \langle v_d(t)\rangle_{T_s} \\ \langle v_{OB}(t)\rangle_{T_s} \end{bmatrix} \tag{14}$$

$$\langle i_B(t)\rangle_{T_s} = \begin{bmatrix} \left(1 - \frac{r_B || r_C}{r_C \cdot C}\right) & \frac{1}{r_B} \end{bmatrix} \cdot \begin{bmatrix} \langle i_L(t)\rangle_{T_s} \\ \langle v_C(t)\rangle_{T_s} \end{bmatrix} - \frac{1}{r_B}\cdot \langle v_{OB}(t)\rangle_{T_s}$$

Finally, the model in (14) is linearized using the Taylor series and the perturbation of the variables around the points of the charger [13,20] in steady state. The equilibrium model of the charger, as represented in (12) [17,19], is used to calculate the steady state values. It should be noted that in the switched model of the charger described in (13), the switching function is replaced by its averaged version, with the addition of a small disturbance in ac, i.e. $d(t)$. The linearization of the charger results in the derivation of the small-signal state-space model, as shown as follows:

$$\begin{cases} \dot{\mathbf{x}}(t) = \mathbf{A_m} \cdot \mathbf{x}(t) + \mathbf{B_m} \cdot \mathbf{u}(t) \\ \mathbf{y}(t) = \mathbf{C_m} \cdot \mathbf{x}(t) + \mathbf{D_m} \cdot \mathbf{u}(t) \end{cases} \tag{15}$$

It should be noted that the variables with hats are the small-signal variables, which are assumed to be much smaller than their steady-state variables [17].The state vector, defined in (15) as $\mathbf{x}(t) = \left[\hat{i}_{\mathrm{L}}(t),\ \hat{v}_{\mathrm{C}}(t)\right]^{\mathrm{T}}$, groups the state variables. The input vector, defined as $\mathbf{u}(t) = \left[\hat{v}_{\mathrm{d}}(t),\ \hat{v}_{\mathrm{OB}}(t),\ \hat{d}(t),\right]^{\mathrm{T}}$, groups the input variables. Finally, the output vector, defined as $\mathbf{y}(t)$ = $[\mathbf{x}(t),\ \hat{i}_{\mathrm{B}}(t)]^{\mathrm{T}}$, groups the input variables. In this charger, the outputs have been defined as the state variables plus the load current. Furthermore, $\mathbf{x}(t) \in \{\mathbb{R}^2\}$ and $\{\mathbf{u}(t), \mathbf{y}(t)\} \in \{\mathbb{R}^3\}$. Conversely, the matrices of the model in (15) are defined as follows:

$$\mathbf{A_m} = \begin{bmatrix} -\dfrac{R_{\mathrm{in}} + r_{\mathrm{B}} \,||\, r_{\mathrm{C}}}{L} & -\dfrac{r_{\mathrm{B}} \,||\, r_{\mathrm{C}}}{r_{\mathrm{C}} \cdot L} \\ \dfrac{r_{\mathrm{B}} \,||\, r_{\mathrm{C}}}{r_{\mathrm{C}} \cdot C} & \dfrac{r_{\mathrm{B}} \,||\, r_{\mathrm{C}}}{r_{\mathrm{C}} \cdot r_{\mathrm{B}} \cdot C} \end{bmatrix}, \mathbf{B_m} = \begin{bmatrix} \dfrac{D}{L} & -\dfrac{r_{\mathrm{B}} \,||\, r_{\mathrm{C}}}{r_{\mathrm{B}} \cdot L} & \dfrac{V_{\mathrm{d}}}{L} \\ 0 & \dfrac{r_{\mathrm{B}} \,||\, r_{\mathrm{C}}}{r_{\mathrm{B}} \cdot r_{\mathrm{C}} \cdot C} & 0 \end{bmatrix},$$
$$\mathbf{C_m} = \begin{bmatrix} 1 & 0 \\ 0 & 1 \\ \dfrac{r_{\mathrm{B}} \,||\, r_{\mathrm{C}}}{r_{\mathrm{B}}} & \dfrac{r_{\mathrm{B}} \,||\, r_{\mathrm{C}}}{r_{\mathrm{B}} \cdot r_{\mathrm{C}}} \end{bmatrix}, \mathbf{D_m} = \begin{bmatrix} 0 & 0 & 0 \\ 0 & 0 & 0 \\ 0 & -\dfrac{r_{\mathrm{B}} \,||\, r_{\mathrm{C}}}{r_{\mathrm{B}} \cdot r_{\mathrm{C}}} & 0 \end{bmatrix} \tag{16}$$

From here, $\mathbf{A_m}$, $\mathbf{B_m}$, $\mathbf{C_m}$, and $\mathbf{D_m}$ are the state, input, output and direct transmission matrices respectively. Also, $\mathbf{A_m}\ \mathcal{M}_{2\mathrm{x}2} \in \{\mathrm{K}\}$, $\mathbf{B_m}\ \mathcal{M}_{2\mathrm{x}3} \in \{\mathrm{K}\}$, $\mathbf{C_m}$, $\mathcal{M}_{3\mathrm{x}2} \in \{\mathrm{K}\}$, and $\mathbf{D_m}\ \mathcal{M}_{3\mathrm{x}3} \in \{\mathrm{K}\}$.

## 4. Efficiency calculation

An interesting figure of merit to characterize the charger is the expression of the performance parameter ($\eta$) [17,21–23]. It is known that $\eta = P_{\mathrm{o}}/P_{\mathrm{i}}$, where $P_{\mathrm{i}}$ and $P_{\mathrm{o}}$ are the input and output power of the charger, respectively. According to **Figure 3**, $P_{\mathrm{i}}$ and $P_{\mathrm{o}}$ are given by (17). From (17), $A_{\mathrm{v1}}$ and $A_{\mathrm{v2}}$ are defined as steady-state voltage gains, specifically $A_{\mathrm{v1}} = V_{\mathrm{C}}/V_{\mathrm{d}}$ and $A_{\mathrm{v2}} = V_{\mathrm{C}}/V_{\mathrm{OB}}$. Then, the expression of $\eta$ is calculated and shown in (18).

$$\begin{cases} P_{\mathrm{i}} = D \cdot \left( \dfrac{A_{\mathrm{V}_1}}{r_{\mathrm{B}} \,||\, R_{\mathrm{in}}} - \dfrac{D}{R_{\mathrm{in}}} \right) \cdot \left( A_{\mathrm{V}_2} - 1 \right) \cdot V_{\mathrm{d}}^{2} \\ P_{\mathrm{o}} = r_{\mathrm{B}} \cdot \left( \dfrac{A_{\mathrm{V}_1}}{r_{\mathrm{B}} \,||\, R_{\mathrm{in}}} - \dfrac{D}{R_{\mathrm{in}}} \right)^{2} \cdot V_{\mathrm{d}}^{2} \end{cases} \tag{17}$$

$$\eta = r_{\mathrm{B}} \cdot \left( \frac{A_{\mathrm{V}_1}}{r_{\mathrm{B}} \,||\, R_{\mathrm{in}}} - \frac{D}{R_{\mathrm{in}}} \right) \cdot \left( \frac{1}{\left( A_{\mathrm{V}_2} - 1 \right) \cdot D} \right) \tag{18}$$

It is interesting to note that this expression is of the nonlinear type, mainly due to the presence of $A_{\mathrm{v1}}$ and $A_{\mathrm{v2}}$. Furthermore, it can be seen that the efficiency is strongly dependent on the nonlinearities $R_{\mathrm{DS(on)}}$ and $r_{\mathrm{L}}$, and it can be confirmed from (18) that this dependence is of the hyperbolic type.

## 5. *L* and *C* sizing

In order to calculate the energy storage elements, $L$ and $C$, the slopes ($m$) of $i_{\mathrm{L}}(t)$ and $v_{\mathrm{C}}(t)$ during the first switching half-cycle, i.e., in the time interval $0 \leq t < D{\cdot}T_{\mathrm{s}}$, are considered. These slopes are obtained from (4) and (5), respectively. Additionally, the slopes of $i_{\mathrm{L}}(t)$ and $v_{\mathrm{C}}(t)$ are depicted in **Figure 4(a)** and **(b)**, respectively. Based on **Figure 4** and the aforementioned equations, the values of $L$ and $C$ are calculated and presented as follows:

$$L \geq \frac{1}{2} \cdot \frac{D \cdot T_{\mathrm{s}}}{\Delta I_{\mathrm{L}}} \cdot \left( V_{\mathrm{d}} - \left( \frac{r_{\mathrm{B}} + R_{\mathrm{in}}}{r_{\mathrm{B}}} \right) \cdot V_{\mathrm{C}} + \frac{R_{\mathrm{in}}}{r_{\mathrm{B}}} \cdot V_{\mathrm{OB}} \right) \tag{19}$$

$$C \geq \frac{1}{2} \cdot \frac{D \cdot T_{\mathrm{s}} \cdot r_{\mathrm{B}} \,||\, r_{\mathrm{C}}}{r_{\mathrm{C}} \cdot \Delta V_{\mathrm{C}}} \cdot \left( I_{\mathrm{L}} - \frac{1}{r_{\mathrm{B}}} \cdot V_{\mathrm{C}} + \frac{1}{r_{\mathrm{B}}} \cdot V_{\mathrm{OB}} \right) \tag{20}$$

From (19) and (20), the maximum allowable ripple percentages with respect to $V_{\mathrm{C}}$ and $I_{\mathrm{L}}$, i.e., $\Delta V_{\mathrm{C}}$ and $\Delta I_{\mathrm{L}}$ respectively, can be calculated. It can be seen also, that both $\Delta V_{\mathrm{C}}$ and $\Delta I_{\mathrm{L}}$ represent the constraints imposed on 2L-BC, which must be met. It is once again evident that the nonlinearities of the charger affect the sizing of the energy storage elements, specifically at $L$ and $C$. Nevertheless, as indicated by (19) and (20), the sizing of $L$ is more significantly influenced by the variation of $R_{\mathrm{DS(on)}}$ and $r_{\mathrm{L}}$ than by the sizing of $C$.

## 6. Compensator synthesis

The linear model in (15) is considered, and the Laplace transform ($s$) is applied to obtain the $s$-domain model shown as follows:

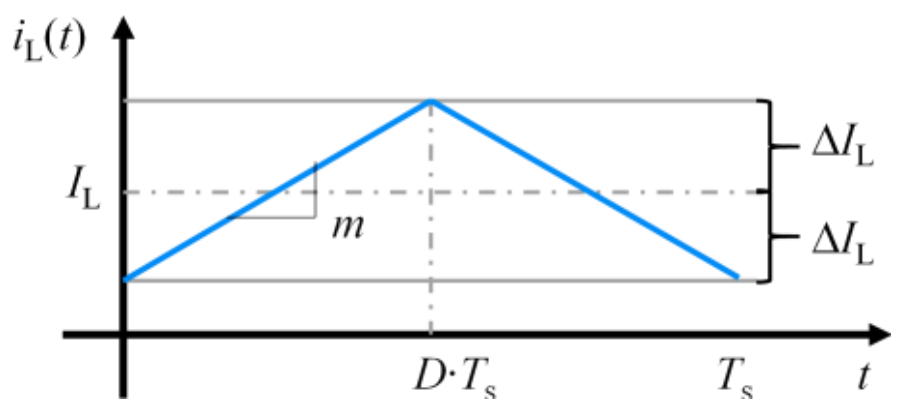

(a)

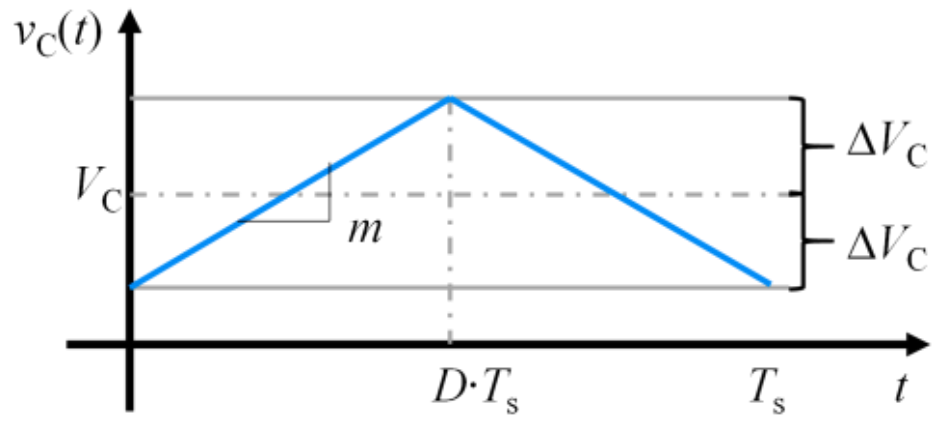

(b)

**Figure 4.** Waveforms of $i_L(t)$ and $v_C(t)$ in the steady state of 2L-BC. (a) $i_L(t)$ waveform. (b) $v_C(t)$ waveform.

$$\mathbf{Y}(s) = \left[\mathbf{C_m} \cdot (s \cdot \mathbf{I} - \mathbf{A_m})^{-1} \cdot \mathbf{B_m} + \mathbf{D_m}\right] \cdot \mathbf{U}(s) \tag{21}$$

In (21), **I** is an identity matrix of dimension 2x2. From (21), it follows that $\mathbf{Y}(s) = [I_L(s), V_C(s), I_B(s)]^T$ and $\mathbf{U}(s) = [V_d(s), V_{OB}(s), D(s)]^T$ are the complex output and input vectors, respectively. These vectors are defined as $\{\mathbf{U}(s), \mathbf{Y}(s)\} \in \{\mathbb{C}^3\}$.

**Figure 5** depicts the proposed control diagram for the 2L-BC, whose topology is based on a standard output feedback structure [20]. The primary control objective is to regulate $i_B(t)$ through a PI compensator by providing the controlled variable $d(t)$. The variable $d(t)$ is the duty ratio as a function of time. Subsequently, $d(t)$ is supplied to the charger modulator, which provides the switching function $s_f(t)$. This function relates the states of switches $Q_1$ and $Q_2$ in accordance with (1).

The transfer function of interest for this control system is obtained from (21), which corresponds to:

$$\left.\frac{I_B(s)}{D(s)}\right|_{\substack{V_d(s)=0 \\ V_{OB}(s)=0}} = k \cdot \frac{\left(1 + \dfrac{s}{\omega_z}\right)}{\left(1 + \dfrac{s}{\omega_{p_1}}\right) \cdot \left(1 + \dfrac{s}{\omega_{p_2}}\right)} \tag{22}$$

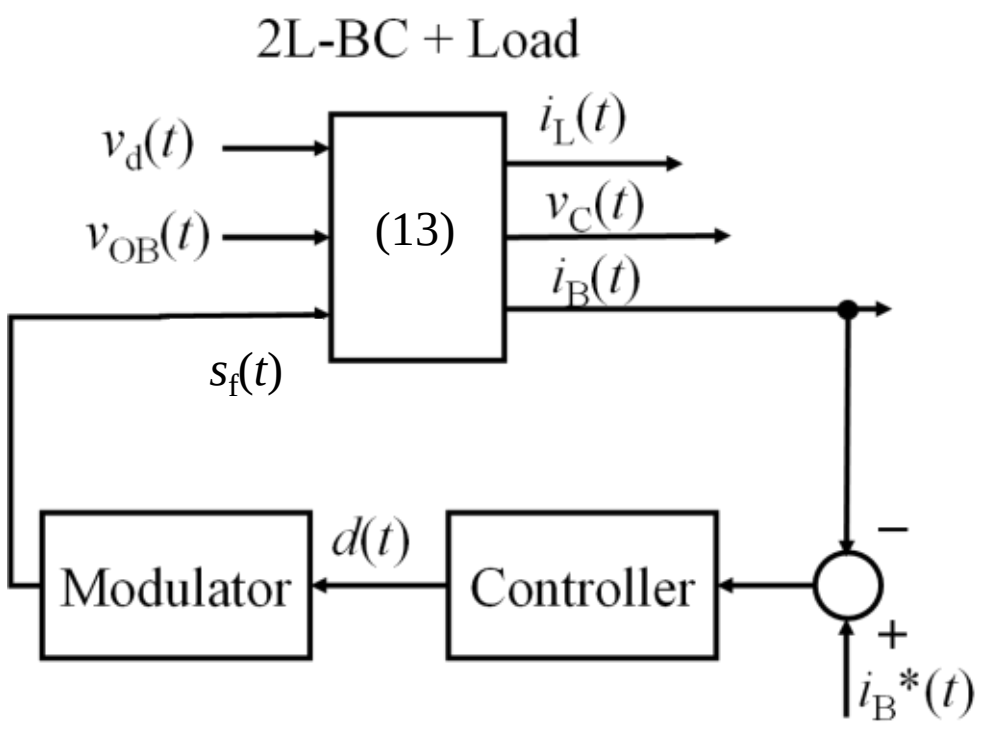

**Figure 5.** Proposed control system for the 2L-BC. The proposed control diagram is based on the standard feedback structure, based on a PI feedback compensator and the linear charger plant defined by (15). It should be noted that, in this diagram, the switched and nonlinear dynamics of the converter and the battery are present in the same block labelled by (13).

The steady-state gain ($k$), the zero ($\omega_z$), and the poles ($\omega_{p1}$ and $\omega_{p2}$) are defined as follows:

$$k = \frac{V_d}{L \cdot C \cdot (r_B + r_C) \cdot \omega_{p_1} \cdot \omega_{p_2}}$$

$$\omega_{p_1} = -0.5 \cdot R_{eff\,1} + \sqrt{(0.5 \cdot R_{eff\,1})^2 - R_{eff\,2}}$$

$$\omega_{p_2} = -0.5 \cdot R_{eff\,1} - \sqrt{(0.5 \cdot R_{eff\,1})^2 - R_{eff\,2}} \tag{23}$$

$$R_{eff\,1} = \frac{r_B \,||\, r_C}{C \cdot r_B \cdot r_C} + I_L \cdot (R_{in} + r_B \,||\, r_C)$$

$$R_{eff\,2} = \frac{(r_B \,||\, r_C)^2}{C \cdot r_C} \cdot \left( \frac{1}{L \cdot r_C} + \frac{I_L}{r_B} \right) + \frac{I_L \cdot R_{in} \cdot r_B \,||\, r_C}{C \cdot r_C \cdot r_B}$$

The controller described in **Figure 5** is designed using the phase-margin test (PMT) [17]. As illustrated in **Figure 5**, and in accordance with the extensive literature on PI compensators, this compensator can be modeled as follows [17]:

$$G_c(s) = k_p \cdot (1 + 1/\tau_i/s) \tag{24}$$

where $k_p$ and $\tau_i$ are the proportional and integral constant respectively. The block diagram associated with this loop is illustrated in **Figure 6**. On the other hand, the loop gain of the system is defined in (25).

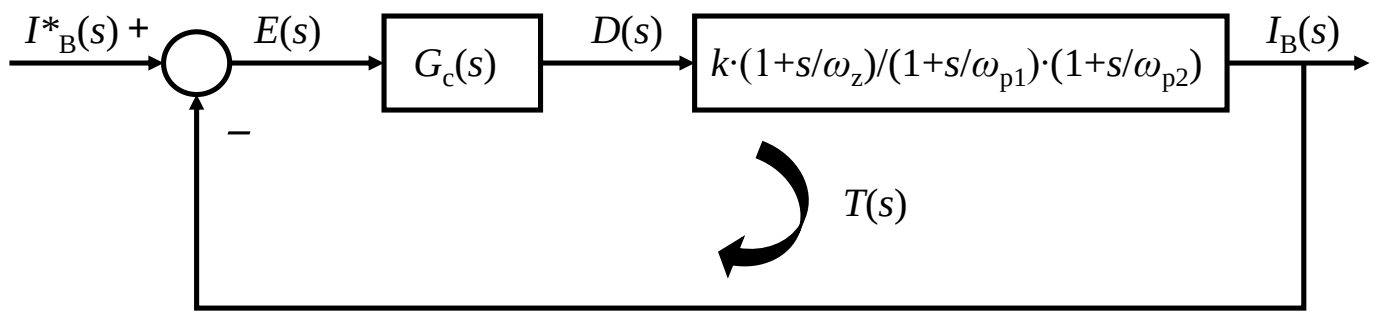

**Figure 6.** Block diagram of the charger charging current control loop.

$$T(s)=G_c(s)\cdot k\cdot\frac{\left(1+\dfrac{s}{\omega_z}\right)}{\left(1+\dfrac{s}{\omega_{p_1}}\right)\cdot\left(1+\dfrac{s}{\omega_{p_2}}\right)} \tag{25}$$

On the other hand, the modulator it is based on pulsewidth modulation (PWM) and its design and operation are described in **Figure 7** and the following [17,18]. Its Laplace model is given by $M(s) = 1/V_M$, where, in this case, $V_M = 1$ V. The high-frequency carry function is represented by the following model.

$$s_f(t)=\begin{cases}1, & d(t)>v_{carry}(t)\\ 0, & d(t)<v_{carry}(t)\end{cases} \tag{26}$$

According to the PMT, it is necessary to study the frequency behavior of the loop gain where the compensator $G_c(s)$ is not involved, assuming that $G_c(s) \approx 1$. This loop gain ($T_u(s)$) is given by:

$$T_u(s)=k\cdot\frac{\left(1+\dfrac{s}{\omega_z}\right)}{\left(1+\dfrac{s}{\omega_{p_1}}\right)\cdot\left(1+\dfrac{s}{\omega_{p_2}}\right)} \tag{27}$$

Table 1 provides a list of the parameters associated with the charger under study. Table 1 indicates that the frequency response of (27) can be obtained and illustrated in **Figure 8**. With regard to the PMT, the crossover frequency ($f_c$) of the frequency response of $T_u(s)$ can be considered to be equal to the switching frequency of the charger, that is, $f_c = f_s = 27$ kHz. Consequently, **Figure 8** reveals that $f_c \approx 13.2$ kHz. Consequently, a compensator is necessary to bring the $f_c$ of the uncompensated system as close as possible to $f_s$. As part of the procedure, the compensated loop gain of the system has been defined in (25). However, by making the change of variable $s = j\cdot\omega$, assuming $\omega >> 1$, defining that $\omega_c = \omega_s$, and taking into account (24) one has:

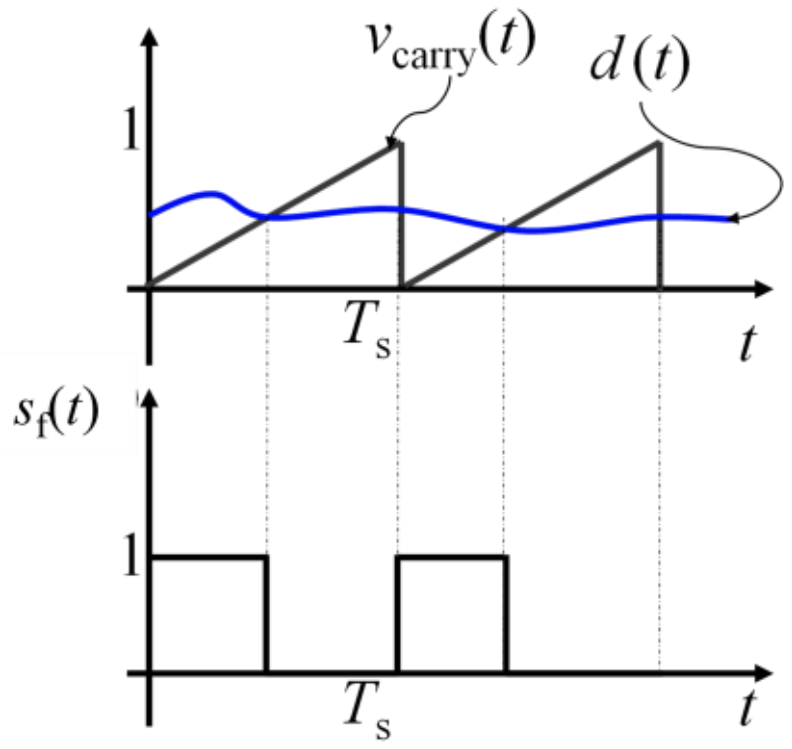

**Figure 7.** PWM generation for the 2L-BC. A sawtooth waveform is compared with the signal $d(t)$ and as a result of this operation, according to (19), the switching function $s_f(t)$ is generated.

**Table 1.** 2L-BC parameters.

| Parameter | Value |
|---|---|
| $R_{DS(on)}$ | 35 [mΩ] |
| $r_L$ | 1 [Ω] |
| $r_C$ | 1.5 [Ω] |
| $r_C$ | 1 [Ω] |
| $L$ | 9.5 [mH] |
| $C$ | 100 [ηF] |
| $v_{OB}$ | 450 [V] |
| $D$ | 0.9 |
| $\Delta V_C$ | 5 [%] |
| $\Delta I_L$ | 5 [%] |

$$T(\mathrm{j}\cdot\omega) = -\mathrm{j}\cdot\frac{k\cdot k_p}{\omega_c} \tag{28}$$

valid for high frequencies. The PMT can be applied to yield the following result in (29) [17], and, to ensure good stability, $\tau_i = 100/\omega_s$ is defined [17]. Consequently, the PI compensator has already been designed. In order to verify that the PI has been properly designed, (25) must be evaluated. **Figure 9** illustrates the frequency response of the compensated loop gain.

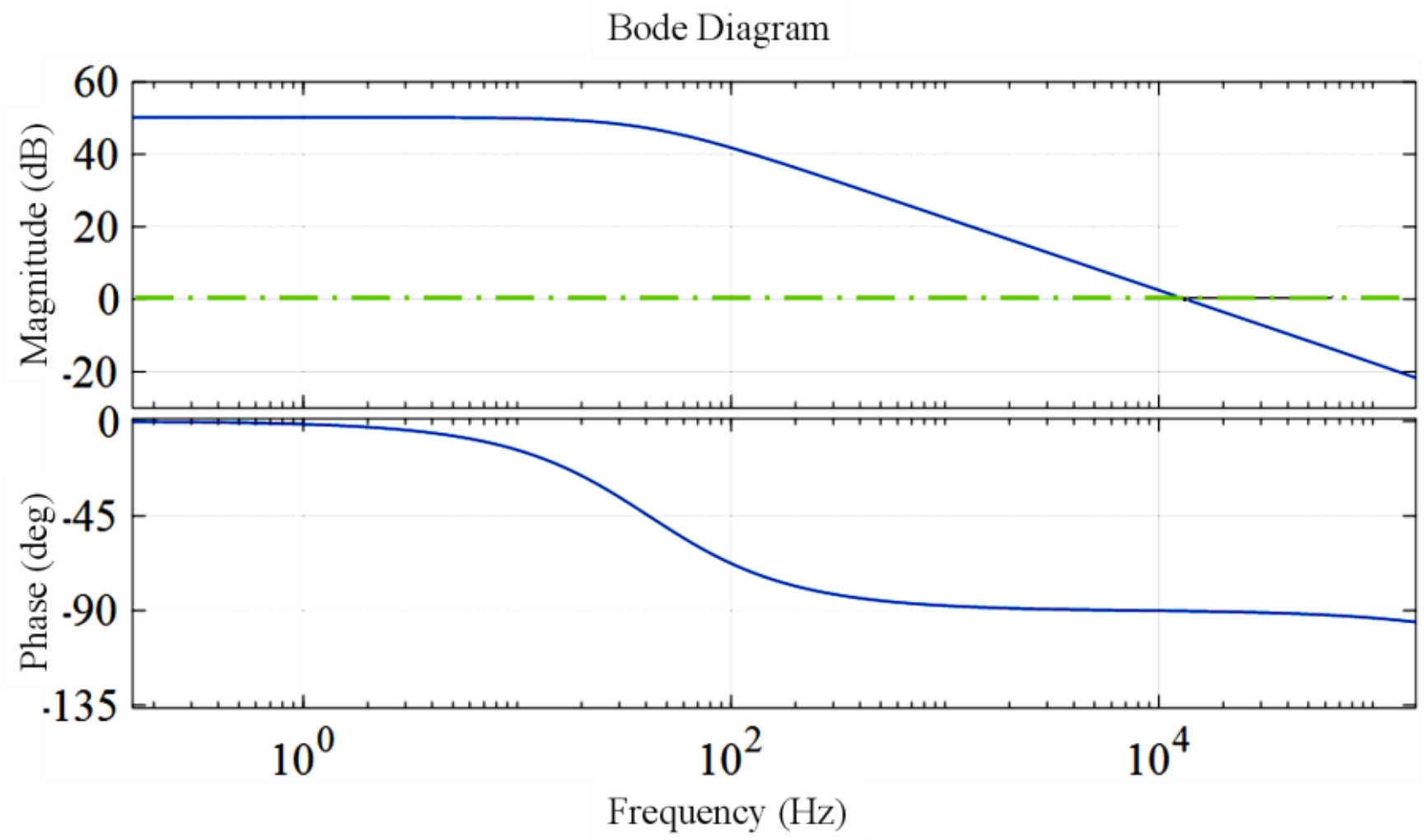

**Figure 8.** Frequency response of (27). $f_c \approx 13.2$ kHz.

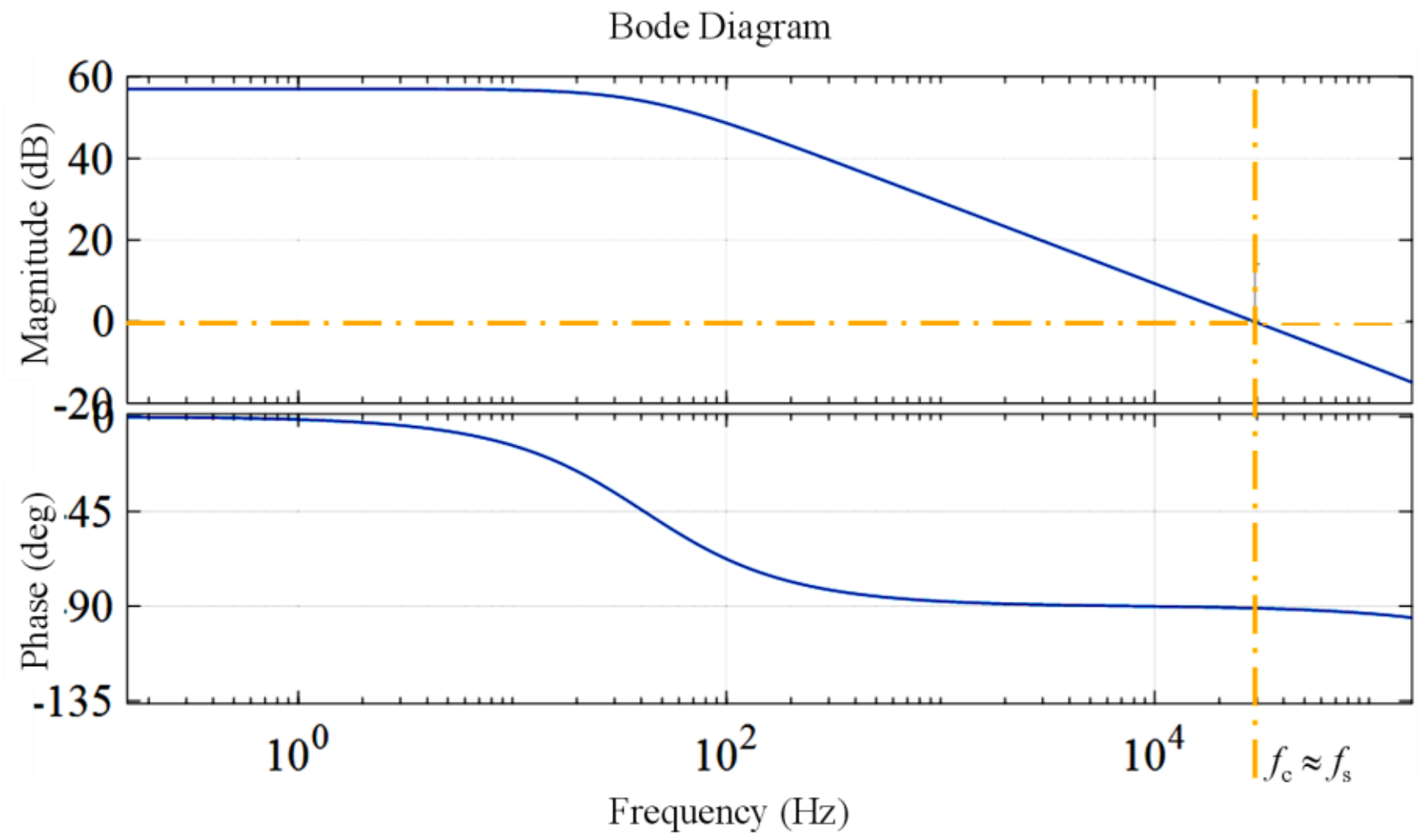

**Figure 9.** Frequency response of (25). $f_c \approx 29$ kHz.

$$\text{abs}\left( T(\text{j} \cdot \omega) \Big|_{\substack{\text{@high frequency} \\ \omega = \omega_c}} \right) = 1 \Rightarrow \frac{k \cdot k_p}{\omega_c} = 1 \Rightarrow k_p = \frac{\omega_s}{k} \Big|_{\omega_s = \omega_c} \tag{29}$$

From **Figure 9**, it can be established that the $f_c$ (≈ 29 kHz) is close to $f_s$, which indicates that the design of the PI compensator is adequate.

## 7. Simulation results

The simulation of the switched and averaged 2L-BC models defined in (13) and (14), respectively, is performed using MATLAB-Simulink and Table 1. The control system proposed in **Figure 5** is also implemented. The 2L-BC is supplied with a dc voltage $V_d$ = 800 V and operates with an $f_s$ of 27 kHz.

The initial conditions of the 2L-BC are imposed on the battery and inductor currents, as well as on the capacitor voltage. This is indicated by the following values: $i_{B0} = i_{L0} = 0$ A and $v_{C0} = 400$ V. In contrast, the initial reference value of the load current is $I^*_B = 30$ A.

**Figure 10** depicts the simulation results for the transient dynamic responses of $i_B(t)$, $i_L(t)$, $v_C(t)$, $d(t)$, and $v_{OB}(t)$ in their switched version, i.e., $i_{Bsm}(t)$, $i_{Lsm}(t)$, $v_{Csm}(t)$, $d_{sm}(t)$, $v_{OBsm}(t)$, and in their averaged version, i.e., $i_{Bav}(t)$, $i_{Lav}(t)$, $v_{Cav}(t)$, $d_{av}(t)$, and $v_{OBav}(t)$, respectively. Regarding the 2L-BC operation, at 60 ms, there is a step change in $i_B(t)$, reaching a new value of 40 A and subsequently entering a steady state. Additionally, at 90 ms, a change in the load is generated, resulting in a shift in the internal battery voltage, $v_{OB}(t)$, from 450 V to 350 V. This disturbance in the system dynamics is observed at 90 ms.

**Figure 10(a)** illustrates that $i_B(t)$ behaves as if it were a first-order system, exhibiting no overshoot and zero steady-state errors. The settling times are notably brief, with $i_B(t)$ reaching its steady state after startup in just 30 ms. Following the step change in $i^*_B(t)$, $i_B(t)$ reaches its final steady-state value in 4 ms. Furthermore, upon the occurrence of a change in load, a transient is generated in $i_B(t)$ of a very short duration (resembling a pulse, as described by [20]), with a duration of approximately 1.5 µs, after which it returns to its steady state value of 40 A. This situation serves to illustrate the optimal design of the PI compensator, while also demonstrating the adequate stability of the system.

Conversely, it can be demonstrated that the sizing of $L$ and $C$ is adequate and meets the ripple requirement. That is, the ripple of $i_L(t)$ and $v_C(t)$ is lower than $\Delta I_L$ and $\Delta V_C$, respectively.

As illustrated in **Figure 10(b)** and **(c)**, it can be observed that the values of $\Delta I_L$ and $\Delta V_C$ reach 0.16% and 2.4%, respectively.

In addition, the dynamics of $i_L(t)$ and $v_C(t)$ are observed to work correctly with a step change in $i^*_B(t)$ and a change in load. After starting, $v_C(t)$ takes the value of 480 V and $i_L(t)$ the value of 30 A. Then, for the step change in $i_B(t)$, a logical increase in both variables (in $i_L(t)$ and $v_C(t)$ respectively) is generated. Finally, when the disturbance is generated in the load, $i_L(t)$ takes its maximum value of 45.43 A and after 13 ms it returns to its value of 40 A. On the other hand, $v_C(t)$ suffers a significant decrease, taking a transient value of 395.4 V and after 10 ms it reaches its final steady state value of 390 V. Such a condition generated in both variables after the load disturbance can be verified by analyzing expressions (2), (3), (6) and (7).

In **Figure 10(d)**, it is observed that after the system startup, the duty ratio $d(t)$ takes the value of 0.64. After the step change in $i_B(t)$, $d(t)$ takes a new steady state value of 0.66 and finally, when the load change occurs, $d(t)$ undergoes an impulse type disturbance and reaches its minimum value of 0.01 for a time of 1.5 µs (similar to $i_B(t)$). It also assumes a new steady state value of 0.54. On the other hand, a stable and unsaturated behavior is observed at $d(t)$, which implies a greater slack in the control of $i_B(t)$. In addition, a small overshoot in d is observed in response to the step change in $i_B(t)$. It is also observed that the switching frequency $f_s$ is transferred to $d(t)$ due to the operation of the switches according to the switching strategy described in (20).

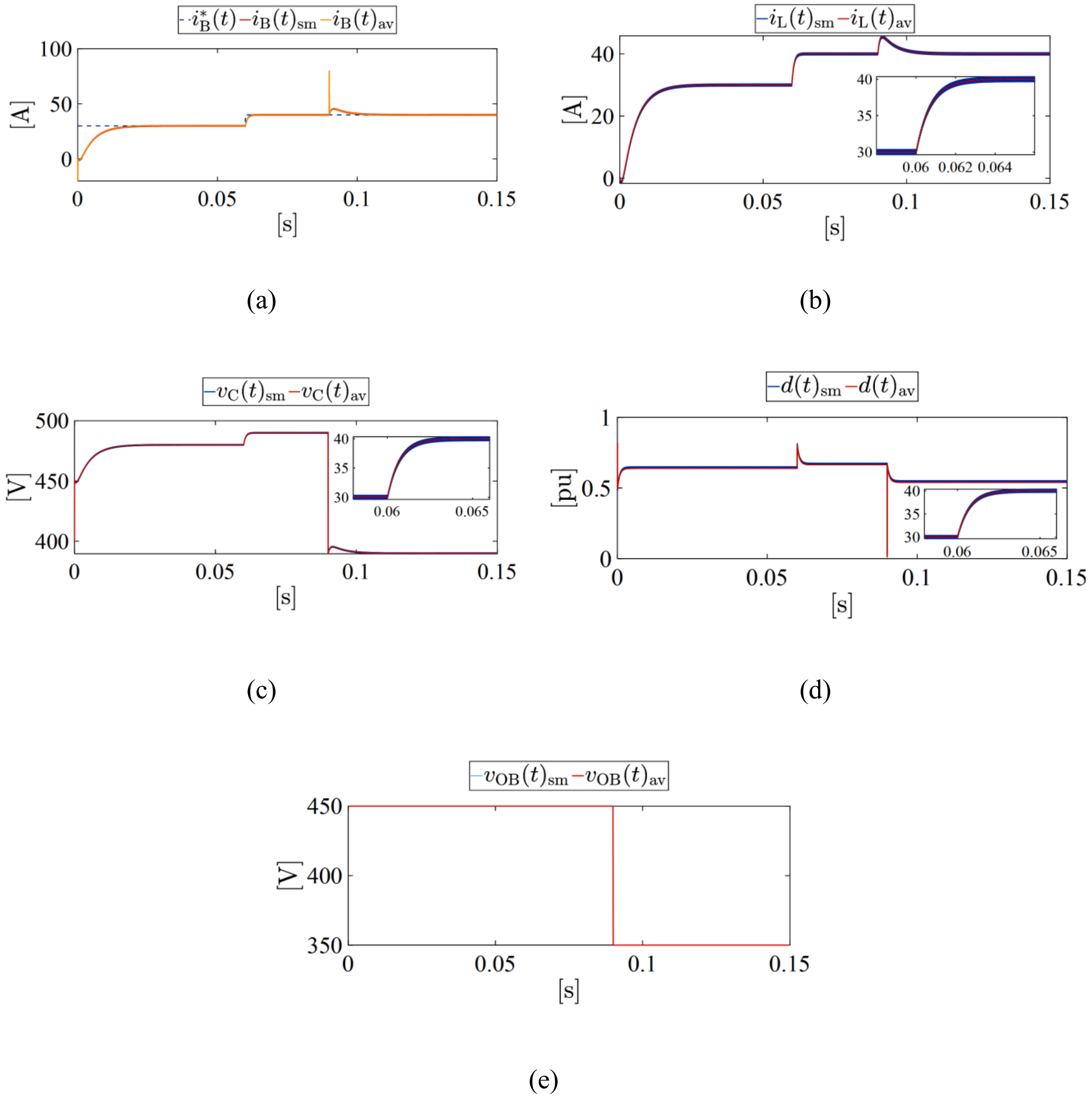

**Figure 10.** Simulation results of the 2L-BC in transient operation with initial conditions of $i_{B0}$ = $i_{L0}$ = 0 A and $v_{C0}$ = 400 V. Step change in $i_B(t)$ at 60 ms and a disturbance in $v_{OB}(t)$ at 90 ms. (a) Dynamic response of $i_B(t)$. (b) Dynamic response of $i_L(t)$. (c) Dynamic response of $v_C(t)$. (d) Dynamic response of $d(t)$. (e) Disturbance in $v_{OB}(t)$.

**Figure 11** depicts a series of surfaces where the focus of analysis is the variation of the equilibrium point $V_C$ as a function of $r_L$ and $R_{DS(on)}$ (see **Figure 11(a)**), the variation of the magnitude of $L$ as a function of $r_L$ and $R_{DS(on)}$ (see **Figure 11(b)**), and the variation of the magnitude of $C$ as a function of $r_L$ and $r_C$ (see **Figure 11(c)**).

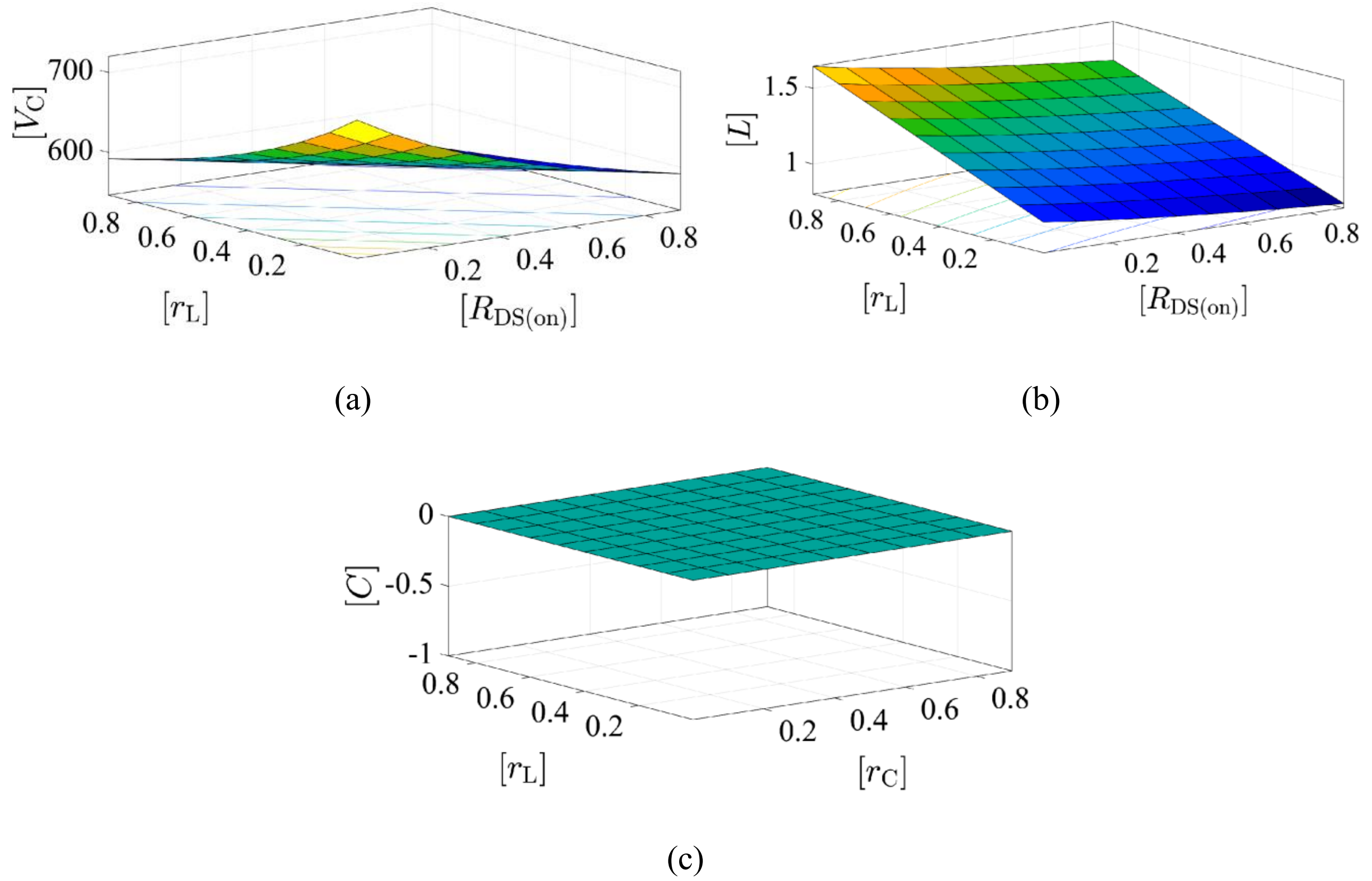

**Figure 11.** 3D plots showing the variations of $V_C$, $L$, and $C$ as a function of $R_{DS(on)}$, $r_L$, and $r_C$. (a) Variation of $V_C$ as a function of $r_L$ and $R_{DS(on)}$. (b) Variation of $L$ as a function of $r_L$ and $R_{DS(on)}$. (c) Variation of $C$ as a function of $r_L$ and $r_C$.

Table 2 presents the values of the parameters that were constant and the respective variables.

For **Figure 11(a)**, the expression associated with $V_C$ is taken from (12). For the case of **Figure 11(b)** (variation of $L$) and (c) (variation of $C$), the expressions associated with them are (19) and (20) respectively.

From **Figure 11(a)** it is clear that $V_C$ decreases significantly as both $r_L$ and $R_{DS(on)}$ increase, i.e. there is an inverse relationship. On the other hand, **Figure 11(b)** shows that $L$ decreases in value as both $r_L$ and $R_{DS(on)}$ decrease, i.e. there is a direct relationship. Finally, from **Figure 11(c)** it can be seen that the magnitude of $C$ remains almost constant regardless of the variations in $r_L$ and $r_C$. This fact allows to confirm what was explained in Section V. Note also that in this figure the value of $C$ is very close to zero for scaling reasons, since its value is around 100 ηF.

**Figure 12** shows the surface of variation of yield as a function of $r_L$ and $R_{DS(on)}$ regarding (18). From the surface, it can be seen that the charger performance is high when the nonlinearities are very small, under ideal conditions, $\eta$ is maximum when $r_L = R_{DS(on)} = 0$. Then, as the nonlinearities increase in value, the performance decreases abruptly. This condition is very important when designing this type of charger. Table 3 presents the values of the parameters that were constant and the respective variables.

**Table 2.** Parameters and variables for the cases illustrated in **Figure 11**.

| Parameter | $V_C = f(R_{DS(on)}, r_L)$ | $L = f(R_{DS(on)}, r_L)$ | $C = f(r_C, r_L)$ |
|---|---|---|---|
| $V_d$ | 800 [V] | 800 [V] | 800 [V] |
| $r_C$ | 1.5 [Ω] | 1.5 [Ω] | 10–6: 1: 103 [Ω] |
| $r_B$ | 1 [Ω] | 1 [Ω] | 1 [Ω] |
| $V_{OB}$ | 450 [V] | 450 [V] | 450 [V] |
| $D$ | 0.9 | 0.9 | 0.9 |
| $\Delta V_C$ | 0.02 [V] | 0.02 [V] | 0.02 [V] |
| $\Delta I_L$ | 0.14 [A] | 0.14 [A] | 0.14 [A] |
| $I_B$ | 30 [A] | 30 [A] | 30 [A] |
| $V_c$ | - | 400 [V] | 400 [V] |
| $R_{DS(on)}$ | 10–6:1:103 [Ω] | 10–6:1: 103 [Ω] | 35 [mΩ] |
| $r_L$ | 10–6: 1: 103 [Ω] | 10–6: 1: 103 [Ω] | 10–6: 1: 103 [Ω] |

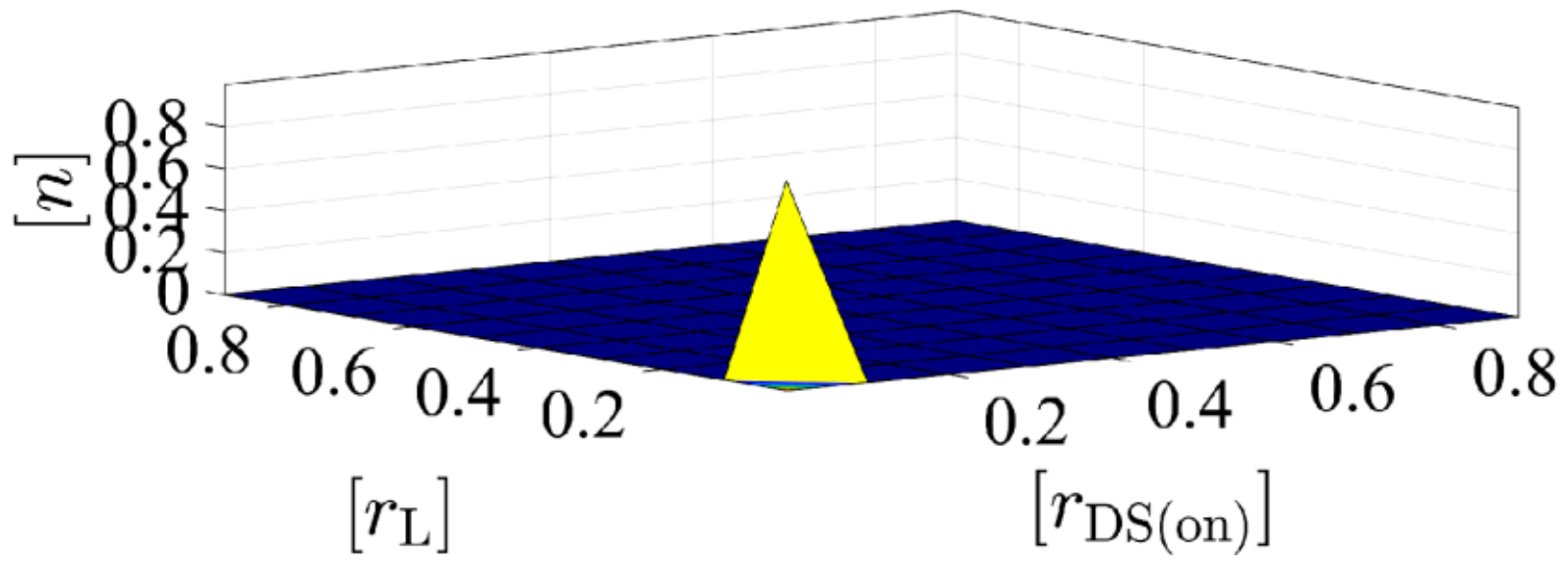

**Figure 12.** 2L-BC performance variation as a function of $R_{DS(on)}$ and $r_L$.

**Figure 13** illustrates the step response of the loop gain in (25). From this figure, it can be observed that the dynamics under transient of (25) behave as a first-order system. Moreover, it can be identified that the system reaches a steady state at approximately 20 ms, demonstrating that the system responds rapidly to disturbances of this nature.

Finally, **Figure 14** depicts the root locus diagram of (25). From this figure, it can be concluded that the loop gain exhibits a totally stable behavior, as all its poles are in the left half-plane of the s-plane. Furthermore, it is demonstrated that the system in question (25) has two zeros and two poles.

**Table 3.** Parameters and variables for variation.

| Case | Parameter | Value |
|---|---|---|
| $\eta = f(R_{DS(on)}, r_L)$ | $V_d$ | 800 [V] |
| | $r_C$ | 1.5 [Ω] |
| | $r_B$ | 1 [Ω] |
| | $V_{OB}$ | 450 [V] |
| | $D$ | 0.9 |
| | $\Delta V_C$ | 0.02 [V] |
| | $\Delta I_L$ | 0.14 [A] |
| | $I_B$ | 30 [A] |
| | $R_{DS(on)}$ | 10–6: 1: 103 [Ω] |
| | $r_L$ | 10–6: 1: 103 [Ω] |

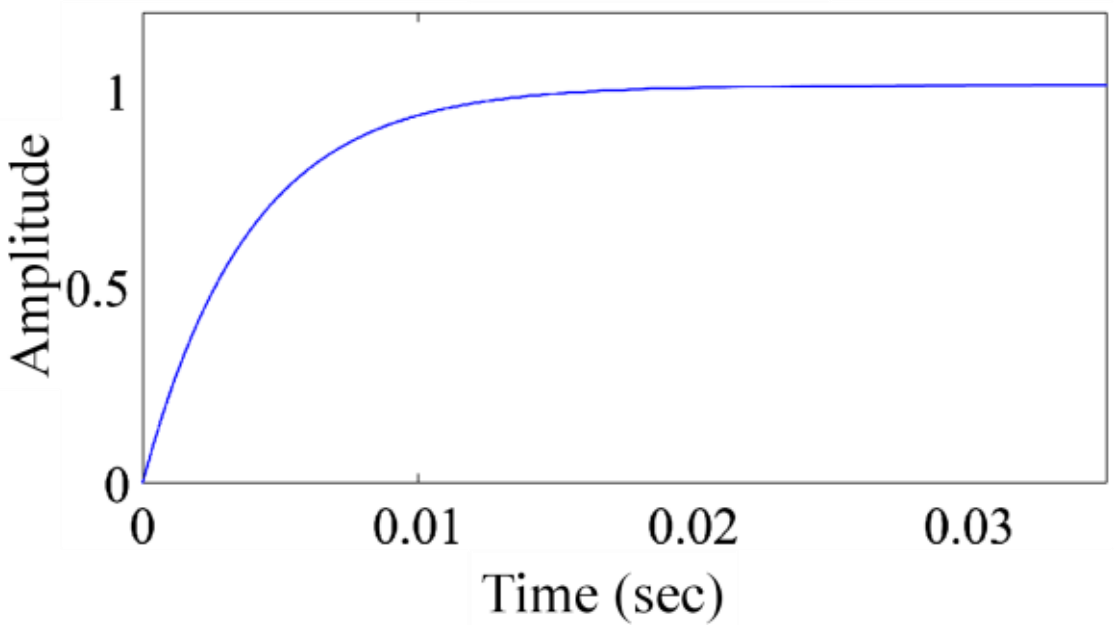

**Figure 13.** Step response of (25).

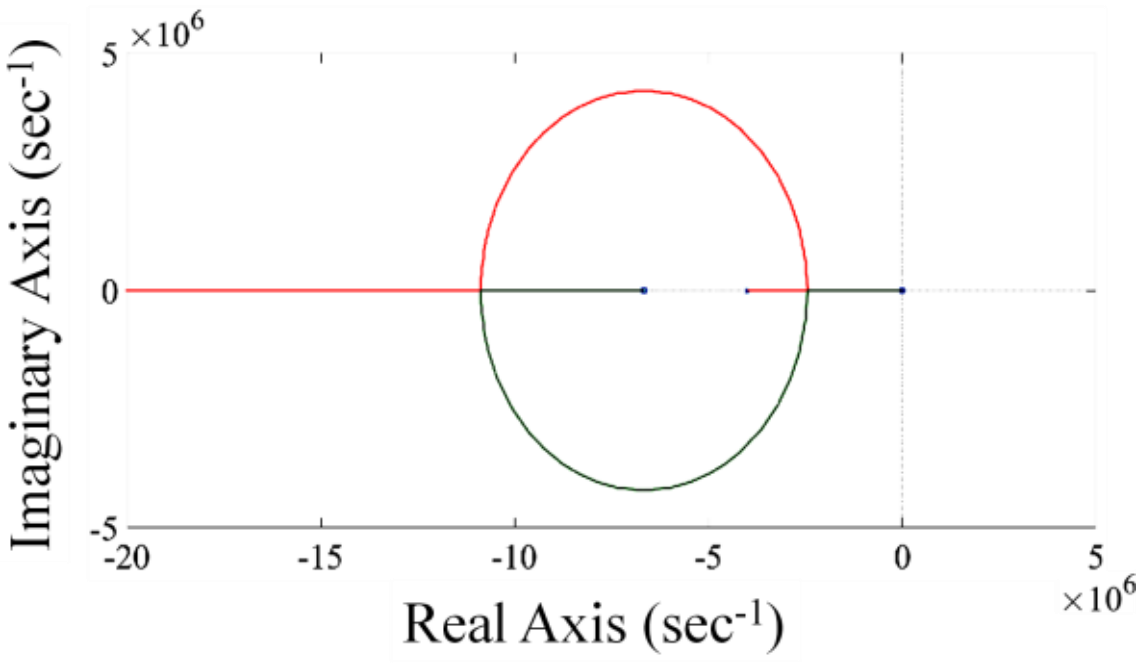

**Figure 14.** Root locus diagram of (25).

## 8. Conclusion

This study presents a comprehensive analysis of a non-ideal two-level battery charger, with a focus on the modeling, simulation, and performance evaluation of the system. The investigation began with an overview of the importance of electric vehicle adoption in mitigating climate change and the critical role of battery chargers in optimizing electric vehicles performance and efficiency.

The proposed two-level battery charger topology was subjected to a comprehensive examination, with particular emphasis placed on its configuration with two switches operating in a complementary manner, in addition to the inclusion of key components such as inductors, capacitors, and batteries. Through meticulous modeling in both steady-state and dynamic regimes, the behavior of the charger under a multitude of operating conditions was meticulously analyzed, taking into account the nonlinearities associated with switch and energy storage element characteristics.

Efficiency calculations revealed a strong dependence of charger performance on nonlinear factors, particularly the resistance of switches and inductors. This highlights the need for precise modeling and control strategies to optimize charger efficiency and performance.

The sizing of energy storage elements, including inductors and capacitors, was investigated to ensure compliance with ripple requirements, with results demonstrating the significant influence of nonlinearities on component sizing.

The synthesis of controllers was addressed, with a focus on designing a robust proportional-integral (PI) compensator to regulate battery charging current effectively. The control system was validated through simulation results, demonstrating rapid transient responses and stable operation under load disturbances. Furthermore, the study explored the impact of nonlinearities on charger performance through surface analysis, revealing the inverse relationship between charger output voltage and switch resistance, as well as the direct relationship between inductor magnitude and switch resistance.

Furthermore, the step response analysis reveals that the system's transient dynamics resemble those of a first-order system, with rapid convergence to steady state observed within a short timeframe. This indicates the system's robustness in responding to disturbances. Moreover, the root locus diagram demonstrates a stable behavior of the loop gain equation, with all poles residing in the left half-plane of the $s$-plane. The presence of two zeros and two poles further characterizes the system's dynamic behavior. These findings collectively contribute to a comprehensive understanding of the charger system's performance and stability characteristics, which is essential for effective design and implementation.

In conclusion, the findings emphasize the necessity of considering non-idealities in battery charger design and control, offering valuable insights for engineers and researchers striving to develop efficient and reliable charging systems for electric vehicles. Potential future research avenues may include further optimization of control strategies, exploration of alternative topologies, and experimental validation of proposed models to enhance the understanding and implementation of non-ideal battery charging technology.